\documentclass[twocolumn,aps,prl,showpacs]{revtex4}
\usepackage{amssymb}
\newcommand{\be}{\begin{equation}}
\newcommand{\ee}{\end{equation}}
\newcommand{\bea}{\begin{eqnarray}}
\newcommand{\eea}{\end{eqnarray}}
\def\n{\noindent}
\begin{document}

\title{Quantum randomness emerging under gravitational nonlinearity}
\author{Tam\'as Geszti}
\affiliation{Department of the Physics of Complex Systems, \\
E{\"o}tv{\"o}s University; H-1117 Budapest, Hungary \\
{e-mail: \tt geszti@galahad.elte.hu}
}

\begin{abstract}
A scenario is outlined for quantum measurement, assuming that
self-sustaining classicality is the consequence of an attractive
gravitational self-interaction acting on massive bodies, and 
randomness arises already in the classical domain. A simple solvable
model is used to demonstrate that small quantum systems influence big 
ones in a mean-field way, offering a natural route to Born's 
probability rule. 
\end{abstract}

\pacs{PACS numbers: 03.65.BZ, 42.50.-p}
\maketitle

The origin of randomness in quantum mechanics is still unclear, after
eighty years since its first recognition by Born\cite{bn}. The so-called
collapse of the wave function\cite{hei}, hardly more than
a name for the moment when randomness emerges, could not be given
a physical identity so far. Von Neumann attached  a schematic picture 
to the name. Assume a measuring apparatus that is capable to 
distinguish two orthogonal states of a microobject by evolving under 
interaction with the object into macroscopically distinguishable
states. However, if the microobject starts from a superposition of the
basis states, a  two-act scenario\cite{n} follows: the composite
system evolves first unitarily into an entangled superposition of 
object and apparatus, then non-unitary evolution - identified 
with wave-function collapse - erases randomly one of the terms of 
the superposition, resulting in a macroscopically definite outcome.
For the simplest case of a  two-state microobject with basis
states $|+\rangle$ and $|-\rangle$, interacting with a one-dimensional
apparatus of center-of-mass (c.o.m.) wave function $\psi(x,t)$, to be 
analyzed below, the unitary stage of evolution reads

\bea
|\Psi\bigr\rangle=\bigl(c_+|+\rangle+c_-|-\rangle\bigr)\psi(x,0)
	\nonumber\\ 
	\Rightarrow c_+|+\rangle \psi_+(x,t) +c_-|-\rangle\psi_-(x,t),
\label{neumann}
\eea

\n $p=|c_+|^2$ and $1-p=|c_-|^2$ being the respective probabilities
that in the subsequent non-unitary stage the system would collapse 
into the first or the second term of the superposition.

 It is implicit that the switch from unitary to collapse dynamics
takes place somewhere on the borderline between quantum and classical
world. Whether the two stages are separated indeed in time, and if so, 
whether the second can be resolved into more detailed dynamics, remain 
unanswered. The situation is summarized in the Copenhagen
interpretation of quantum mechanics, according to which such questions 
are declared illegal, the quantum and classical domains are distinct, 
the former being a mirror world providing dynamics-like rules to
predict statistics of random events in the latter\cite{whz}. It is one 
of the surprising facts about Nature that this interpretation works  
``for all practical purposes'', as observed by John Bell in a tone of 
dissatisfaction\cite{fapp}. 

 The theory of environment-induced decoherence\cite{dec,giu}, although
offering an important clue about the quantum-classical border by
explaining the gradual disappearance of interference between 
more and more different states as we traverse the mesoscopic domain, 
gives no new insight about the nature of collapse. Well tested on 
its own territory, it seems to offer but a vague clue about collapse: 
as we learn - in fair agreement between theory and experiment - to 
keep decoherence at a lower and lower level\cite{decfree}, we seem to 
observe that in such cases the collapse of the wave function
remains also absent: decoherence seems to be a necessary
condition for collapse, although the nature of that control - if
confirmed - remains to be explained. 

A seriously elaborated alternative approach, stochastic reduction 
models\cite{grwd} are a fancy phenomenology postulating two new things 
at a time: a deeper dynamical level with random noise generation, 
and nonlinear quantum dynamics - no independent evidence seems to 
be around in favor of the resulting scheme. Below we argue that 
the theory probably tries to catch too much at one step: randomness 
need not be put in by hand.

The last remark is based on the observation that for a macroscopic apparatus
described by classically chaotic dynamics and coupled to microscopic
subsystems in a mean-field fashion instead of quantum mechanics, Born's
rule\cite{bn} about quantum probabilities would arise in a natural
way: chaotic dynamics transform weak forces into probabilities via 
amplification of small deflections. For short times even linear
quantum mechanics acts on wave packets in a mean-field way as 
exemplified by the theory of so-called weak measurements\cite{wkm}.
Unfortunately, for the long-time phenomenology this fact is of no use:
mean-field behavior is intrinsically nonlinear. 

Starting with the above motivation, we assume that it suffices to
explain the primordial emergence and subsequent persistence of
classicality - the resistance of macroscopic bodies to 
form superpositions; then randomness is generated in the classical
domain  through some kind of chaotic dynamics. The prospects for such 
an explanation have been strongly improved through the recently
discovered  fact that randomness generated in quasi-classical motion 
appears in purely linear quantum dynamics as well\cite{pid}.

The theory of the inflational universe\cite{inf} offers ample support
to the assumption that classical states of the world emerge around
big-bang-like situations as inhomogeneous fluctuations. Taken that as
granted, one has to explain how those classical states resist to the
kind of quantum spreading described by the linear Schr{\"o}dinger equation.

Self-sustaining classicality in the above sense needs nonlinearity in
the form of an attractive self-interaction acting on the wave
function. It can be regarded as a kind of auto-focusing of matter 
waves, eventually caused by a potential whose source is the matter 
density. The usual two-point interaction construct of ordinary quantum 
mechanics is not satisfactory, since it preserves the linearity of the 
equations of motion, with the spreading of wave packets, formation of 
Schr{\"o}dinger cats and all that. On the other hand, {\em mean-field 
approximation} to a two-point attractive interaction is just the thing 
we need to keep wave packets together. There is a trouble though: 
for the usual kind of interactions, mean field is an approximation, 
not the true nature of things.

In all respects, gravity is the natural candidate\cite{qgr}. It is
none of the familiar class of contact interactions, since it is
encoded into the geometry of space-time. There is no evidence against 
{\sl defining} mass density, the source of gravity, as the quantum
mechanical mean value of the operator of mass density, without
additional quantum fluctuations. In that case there is no quantum
fluctuation of the metric either, and gravity acts as a mean field on
all massive objects, including its sources\cite{nocorr}. Such a scheme 
in the limiting case of Newtonian gravity was described in Di{\'o}si
(1984)\cite{di}, who -- following the protocol of Bialynicki-Birula 
and Mycielski\cite{bbm} -- demonstrated the existence of 
soliton-like wave packets stabilized by gravitation, pointing out that 
gravitation is not able to pull distinct wave packets together. 

However, that is not necessary for self-sustaining classicality. We
suggest a scenario in which although Eq.~(\ref{neumann}) remains valid,
the two apparatus wave functions $\psi_+$ and $\psi_-$ remain close to
each other, glued together by gravitation, although the measurement
interaction tends to pull them apart. As a consequence, the double
wave packet, slightly split to balance between opposing forces, moves 
together under the action of the averaged measurement force, weighted by 
$|c_+|^2$ and $|c_-|^2$ in a mean-field way; then entering a frozen 
random environment, undergoes a random choice with the respective 
probabilities $p$ and $1-p$, in agreement with Born's rule.

As demonstrated below in detail, for the above scenario gravitation
has to be strong enough to protect a double wave packet from splitting
under the action of a measuring interaction which is strong enough to
accelerate it as a whole. 

To simplify the analysis, we start from the two-state microobject
interacting with a one-dimensional massive apparatus,  introduced 
in Eq.~(\ref{neumann}). The model Hamiltonian is

\be
\hat H = |+\rangle\langle+|\otimes \hat h_+ 
	+ |-\rangle\langle-|\otimes \hat h_-,
\label{model}
\ee

\n where

\be
\hat h_\pm=-\frac{\hbar^2}{2M}\partial_x^2 + V_{grav}(x,t)
	\mp F_{meas}\cdot x -F_{div}\cdot x
\label{ham}
\ee

\n contains a constant force $\pm F_{meas}$ representing the interaction 
of the microobject in state $|\pm\rangle$ with the apparatus, as well
as a force $F_{div}$ modelling a landscape of random barriers, frozen 
during the act of measurement, diverting the apparatus towards one of 
the possible outcomes of the measurement.

We set out to solve the time-dependent Schr{\"o}dinger equation for the
pure quantum state on the r.h.s. of Eq.~(\ref{neumann}). Because of
$\langle+|-\rangle=0$, the c.o.m density of the apparatus is
a weighted sum of two non-interfering terms:

\be
\varrho(x,t)=p |\psi_+(x,t)|^2+(1-p) |\psi_-(x,t)|^2.
\label{density}
\ee

To calculate the gravitational potential $V_{grav}(x,t)$, we assume that
the apparatus is a homogeneous massive sphere of mass $M$ and radius
$R$, and - what should be checked afterwards - both the widths $x_0$ 
of the c.o.m wave packets and their distance $d$ are kept small:

\be
x_0,~d<<R;
\label{sizebound}
\ee

\n then\cite{di} 

\bea
V_{grav}(x,t)=\frac{M\omega_{grav}^2}{2}\int dy (x-y)^2\varrho(y,t) 
	\nonumber\\ =\frac{M\omega_{grav}^2}{2}x^2
	-M\omega_{grav}^2\overline{x}x
	+\frac{M\omega_{grav}^2}{2}\overline{x^2}. 
\label{gravpot}
\eea

\n Here we have introduced the characteristic frequency of an extended
object oscillating in its own parabolic gravitational potential,

\be
\omega_{grav}^2=G\frac{M}{R^3},
\label{omegrav}
\ee

\n where $G$ is Newton's gravitational constant. For a typical
condensed-matter density $10^4 Kg~m^{-3}$, one obtains $\omega_{grav}
\approx 10^{-3} Hz$. We notice that Eq.~(\ref{gravpot}) furnishes a 
dimensionally correct potential energy that can be plugged into a 
one-dimensional Schr{\"o}dinger equation to obtain valid estimates. 

The basic estimate is this: to sustain classicality of the apparatus,
the double wave packet should remain together during measurement,
therefore its mean splitting should obey $d\approx 
F_{meas}/M\omega_{grav}^2<R$. On the other hand, in order to obtain 
measurement, its displacement as a whole during the measurement 
interaction of duration $\tau_{meas}$ should be large enough: 
$(F_{meas}/M)\tau_{meas}^2\geq l_0,$ where $l_0$ is a displacement 
on the scale of the diverting potential landscape. The combination of
the two requirements gives 

\be
(\omega_{grav}\tau_{meas})^2 > \frac{l_0}{R}.
\label{timestim}
\ee

In view of the estimate of $\omega_{grav}$ given above, this
inequality means that for a measurement of duration $1~s$, random
decision is controlled on the displacement scale of $10^{-6}$ times
the size of the apparatus. If that scale is fixed about $1~nm$, our
estimates require $R \gtrsim 10^{-3}~m$ for an object big enough to survive
conflicting accelerations without splitting into two parts entangled
with different states of a microobject. That is our tentative
criterion for classicality of the apparatus. Although a massive sphere 
dragged around by interaction with a microobject is certainly not the 
realistic model for a particle detector\cite{el}, the above figures give
sufficient comfort to proceed with the analysis.  

Equations ~(\ref{model}, \ref{ham}, \ref{density}, \ref{gravpot})
define a Hamiltonian depending on the modulus of the wavefunction. 
However, as seen from Eq.~(\ref{gravpot}), that dependence is only
through two time-dependent parameters $\overline{ x^{\alpha}(t)}=
\int x^{\alpha}\varrho(x,t)dx~~(\alpha=1,2)$, therefore the
corresponding Schr{\"o}dinger equation, although nonlinear, can be solved
by elementary tools. 

The tool we use is the equation of motion of a coherent state under
the action of a time-dependent force\cite{gard}, having in mind the
time-dependence of the gravitational force proportional to 
$\overline x$ in Eq.~(\ref{gravpot}). The other forces are taken
constant in time\cite{sht}, so we look for a uniformly accelerating 
motion with $\overline x=A+Bt+Ct^2$. Assuming that $\psi_{\pm} 
= |\alpha_{\pm}\rangle$ are coherent states of width $x_0 = 
(\hbar/M\omega_{grav})^{1/2}$ and respective positions
$\overline x_{\pm} = \sqrt{2}~x_0~\Re~\alpha_{\pm}$, from which
averaging gives

\be
\overline x = p\overline x_+ + (1-p)\overline x_-
\label{com}
\ee

\n for the moving c.o.m. coordinate. Applying properly scaled
formulas from Reference \onlinecite{gard}, a closed integral equation 
is obtained for $\overline x(t)$.  It turns out that as expected, the
equation has a smooth, uniformly accelerating solution, the
coefficients $A, B, C$ of which can be evaluated, and one obtains 

\be
\overline x(t) = \overline x(0)+\overline v(0)~t+\frac{\cal F}{2M}~t^2,
\label{smooth}
\ee

\n where

\be
{\cal F}=2~(p-\frac{1}{2})~F_{meas}~+~F_{div}
\label{ftot}
\ee

\n is the sum of the ``meanfieldized'' measurement force caused by
interaction with the microobject, and the frozen random force
diverting the motion of the apparatus towards one of the possible
outcomes, according to  ${\cal F}\gtrless 0$ . If $F_{div}$ is 
uniformly distributed between $-F_{meas}$ and $F_{meas}$, the 
respective probabilities of moving to the right or to the left 
are $p$ and $1-p$, in accordance with Born's rule.

Eq.~(\ref{smooth}) is a solution appropriate to a special class of 
initial conditions. For a general initial condition, oscillations 
around the smooth path appear. We conjecture that the role of 
decoherence in quantum measurement is to continuously damp such 
oscillations. Preliminary numerical results seem to support that 
expectation; details are currently being studied. 

What has been described above is a gravity-based dynamical model for a 
single detector of a binary observable of a microobject. It shows a
route to the dynamical origin of quantum randomness, with no 
creation of macroscopically split superpositions, and accordingly, 
with no necessity of collapse. Once classical states of matter
are present, gravitation forces macroscopic bodies to remain confined, 
and allows them to explore the possibilities classicality offers to 
generate randomness. In that respect this model performs like
stochastic reduction models\cite{grwd} do, without putting in
randomness by hand. 

The present scheme definitely gets out of the Copenhagen philosophy: 
Eq.~(\ref{density}) materializes the Schr{\"o}dinger wave function as
the source of gravity, which precludes the possibility of relegating 
the wave function into the realm of ideas.

In one important respect, however, the present approach is less
efficient than usual quantum mechanics. This is the rationalization 
of detector anti-correlations: the feature that for a single particle 
interacting with two detectors firing on mutually exclusive values of
an observable, only one of the detectors can give a signal. That
feature, in the root of Heisenberg's original collapse idea\cite{hei}, 
and also very seriously discussed by Einstein\cite{ei}, is generally 
believed to hold, although direct experimental tests are not abundant.

In its Neumannian way, ordinary quantum mechanics offers an
explanation\cite{hardy}. In a trivial extension to a two-detector
situation in which $\psi_{\pm}^{(1)}$ means that detector 1 gives a 
signal or not, and  $\psi_{\pm}^{(2)}$ is the same for detector 2, 
Eq.~(\ref{neumann}) is modified like this:

\bea
|\Psi\bigr\rangle
=\bigl(c_+|+\rangle+c_-|-\rangle\bigr)
	\psi^{(1)}(x,0)\psi^{(2)}(x,0)\nonumber\\  \Rightarrow 
	 c_+|+\rangle\psi_+^{(1)}(x,t)\psi_-^{(2)}(x,t)\nonumber\\ 
	+c_-|-\rangle\psi_-^{(1)}(x,t)\psi_+^{(2)}(x,t).
\label{neumann1}
\eea

This equation contains only terms of the expected anti-correlation
property, therefore after collapsing into one of the terms of the
entangled superposition, a classically observable anti-correlation
would appear between the two detectors.

That nice feature is lost in the present scheme, since
$\psi_+^{i}(x,t)$ and $\psi_-^{i}(x,t)$ (for $i=1,2$) are
macroscopically indistinguishable. At the face of our model, the
respective probabilities of the two-detector outcomes $++$, $+-$,
$-+$, $--$ would be $p(1-p)$, $p^2$, $(1-p)^2$, $(1-p)p$. As we all
believe to know, the correct answer is $0$, $p$, $1-p$, $0$.

That very strongly resembles of interference: destructive on some of
the classically apparent possibilities, constructive on others. The
trouble is that I was unable to identify the corresponding scenario of
interference. One of the less obvious possibilities would be a Berry
phase\cite{berry} inducing a vector potential in the two-detector 
configuration space, that would destabilize the $++$ to $--$
diagonal. As a matter of fact, the last term of Eq.~(\ref{gravpot})
gives rise to a fancy phase factor; however, the configuration-space
curl of the resulting vector potential is vanishing: no destabilizing
force arises. Therefore it is left to the Reader as an exercise to
pin down what is interfering with what.

\acknowledgments
Among the many colleagues with whom I had the pleasure to discuss the 
issues exposed in this paper, I would like to mention those who left 
the most significant traces on my thinking: Lajos Di{\'o}si and Anton 
Zeilinger. This work has been partially supported by the Hungarian 
Research Foundation (grant OTKA T 029544).


\begin{thebibliography}{99}
\bibitem{nocorr} Evidence against the scheme would be detecting
gravity-induced quantum correlations, like the ones of coulombic
origin, familiar in quantum chemistry; obviously an experimental 
check is by many orders of magnitude beyond feasibility.
\bibitem{el} In any of the realistic cases, indirect gravitational
confinement is expected to dominate, e.g. electrons of a photodetector
confined in the coulombic cage of a gravitationally confined crystal.
\bibitem{sht} The assumption of constant external forces is not
essential: a short-time expansion for forces and c.o.m. is converted
into a high-frequency asymptotic series for $\alpha_{\pm}(t)$, whose
coefficients can be calculated in succession.
\end{thebibliography}
\end{document}